# Covalently Integrated CNT@rGO for Superior Conductivity and Cycling Stability in Lithium-Ion Batterie

*Junwen Tang[a#], Jingbo Pang[b#], Jie Wang[c#], Huiming Liang[c], Ao Du[a], Long Kuang[a], Xiaoming Cai[d], Ming Qin[a], Cuixia Yan[a], Wu Zhou[b], Jinming Cai[ace]\**

a Faculty of Materials Science and Engineering, Kunming University of Science and Technology, Kunming, Yunnan 650093, PR China

b School of Physical Sciences, University of Chinese Academy of Sciences, Beijing 100049, China

c Guangdong Morion Nanotechnology Co., Ltd, Guangdong 523815, PR China

d Faculty of Mechanical and Electrical Engineering, Kunming University of Science and Technology, Kunming, Yunnan 650500, PR China

e Southwest United Graduate School, Kunming, 650000, PR China

*Corresponding author. Email: j.cai@kust.edu.cn

[#] These authors contributed equally to this work.

## Abstract

The limitations of conventional conductive agents in lithium-ion batteries, such as carbon black and graphite flakes, have driven the search for high-performance alternatives. Carbon nanotubes (CNTs) and graphene offer exceptional conductivity and lower dosage requirements, but face challenges related to high costs and complex fabrication processes. Here, we report a simple and cost-effective one-step chemical vapor deposition (CVD) method for the ultra-high yield growth (7692.31%) of CNTs on a reduced graphene oxide (rGO) substrate, forming a three-dimensional CNT@rGO composites with covalent integration. When employed as a conductive agent for lithium iron phosphate (LiFePO$_4$) cathodes, the CNT@rGO composites significantly enhance rate performance across 1-6C rates, and demonstrate exceptional cycling stability, achieving 96.32% capacity retention after 300 cycles at 1C. The synergistic structure facilitates multiple conductive pathways, minimizes catalyst residue (0.52%), and ensures uniform dispersion, providing an effective and cost-efficient solution for next-generation battery technology. This study lays the foundation for the large-scale application of high-performance carbon conductive agents in battery technology.

## 1. Introduction

The accelerated development of electric vehicles has led to heightened market expectations for lithium-ion batteries, particularly with regard to safety, energy density, and rate performance.[1] [2] Among various cathode materials, LiFePO$_4$ has gained significant attention due to its superior safety and long cycle life, making it a promising candidate for electric vehicle applications. However, the low rate performance caused by its low electronic and ionic conductivity limits its use in high-power applications.[3] [4] Conducting agents are indispensable components that enhance electrode conductivity and stabilize battery performance.[5] Among various conductive materials, single-walled carbon nanotubes (SWCNTs) are regarded as optimal due to their excellent electrical conductivity and mechanical strength. However, their complex preparation process and elevated manufacturing cost hinder their extensive utilization.[6]

In order to resolve the issues associated with SWCNTs, researchers have been exploring potential alternatives. Graphene and multi-walled carbon nanotubes (MWCNTs) have been identified as promising candidates due to their lower preparation cost compared to SWCNTs and their excellent conductivity.[7] [8] [9] However, both materials exhibit challenges related to poor dispersion and a single dimension of electron transport. It is imperative to achieve uniform dispersion of the conductive agent to establish a robust conductive network for the electrode, as this is crucial for effective electron conduction and ensuring the overall performance of the electrode.[10] [11] To address this challenge, researchers have modified the material structure through in situ growth, doping, and compounding. Gupta et al. prepared multilayered PLA/Lithium Iron Phosphate/Carbon Nanotubes microarchitectural cathode materials using a 3D printing technique. The multilayered porous micro-nanostructures significantly improved the ion and electron transport and exhibited remarkable charge/discharge characteristics with a capacity of 62 mAh g$^{-1}$ at a 10C rate.[12] In a separate study, Gao et al. synthesized lithium iron phosphate/carbon nanotube nanocomposites using a combination of in situ microwave plasma chemical vapor deposition and co-precipitation. The open and highly conductive network established by the CNTs contributed to more efficient ion and electron conduction, with a discharge capacity of 126.1, 111.2, 99.5, and 71.3 mAh g$^{-1}$ at 10 C, 20 C, 30 C, and 50 C rates, respectively, at high rates.[13] In the study by Chen et al., lithium iron phosphate was encapsulated with carbon nanotubes and amorphous carbon nanoshells by in situ autocatalysis. The initial capacity was recorded as 136.1 mAh g$^{-1}$ at 1 C, and after 1000 cycles, the capacity was found to be 129.6 mAh g$^{-1}$. This result indicates remarkable long cycling stability.[14]

Existing studies have modified the active materials at the synthesis stage through in situ growth and other means, which can yield superior properties but are complex and costly, making them unsuitable for large-scale use. In this study, a straightforward one-step growth method is employed to achieve uniform dispersion of the catalyst copper nanoparticles, utilizing melamine as a monatomic dispersant. The growth of carbon nanotubes is subsequently initiated after loading on reduced graphene oxide, culminating in the production of CNT@rGO three-dimensional nanoarchitecture materials. This method ensures a substantial yield of carbon nanotubes, amounting to 7652.31%, while maintaining a minimal catalyst residue of only 0.52%. The three-dimensional nanoarchitecture of the material provides good electron transport properties in all three dimensions and facilitates uniform dispersion among the active materials, with an ICE of 99.88% at 0.1C, and higher capacity retention and cycling stability than rGO, MWCNT, and rGO&MWCNT in 1-6C, with a capacity retention of 96.32% after 300 cycles at 1C. When considered in conjunction with the material structure and the corresponding electrochemical performance tests, CNT@rGO, which was uniformly dispersed among LiFePO$_4$ nanoparticles, played a beneficial bridging and structural stabilization role and constructed an effective conductive network.

## 2. Results and Discussion

### 2.1. Materials Preparation and Characterization

CNT@rGO was prepared using a simple two-step method. In the first step, melamine was mixed with CuCl$_2$·2H$_2$O in an ethanol solution and then dried. This process resulted in Cu being dispersed in the carbon-nitrogen (CN) skeleton in the form of monatomic units, thereby constructing the supramolecular structure. In the second step, the mixture was then homogenously mixed with rGO powder and sent to a CVD tube furnace for a one-step heat treatment. As the temperature increased, the melamine began to be thermally condensed, forming Cu-g-$C_3N_4$ with Cu atoms gradually. The formation of Cu-$C_3N_4$ occurred as the temperature continued to rise. Concurrently, g-$C_3N_4$ underwent gradual decomposition, resulting in the uniform dispersion of monatomic copper on the skeleton. This process subsequently led to the release and polymerization of copper in an orderly manner, culminating in the formation of homogeneous nanoparticles of a specific size. Following the introduction of $C_2H_4$ as a carbon source, the growth of carbon nanotubes was initiated, leading to the synthesis of nanomaterials with a three-dimensional architecture of CNT@rGO upon cooling (Figure 1a).

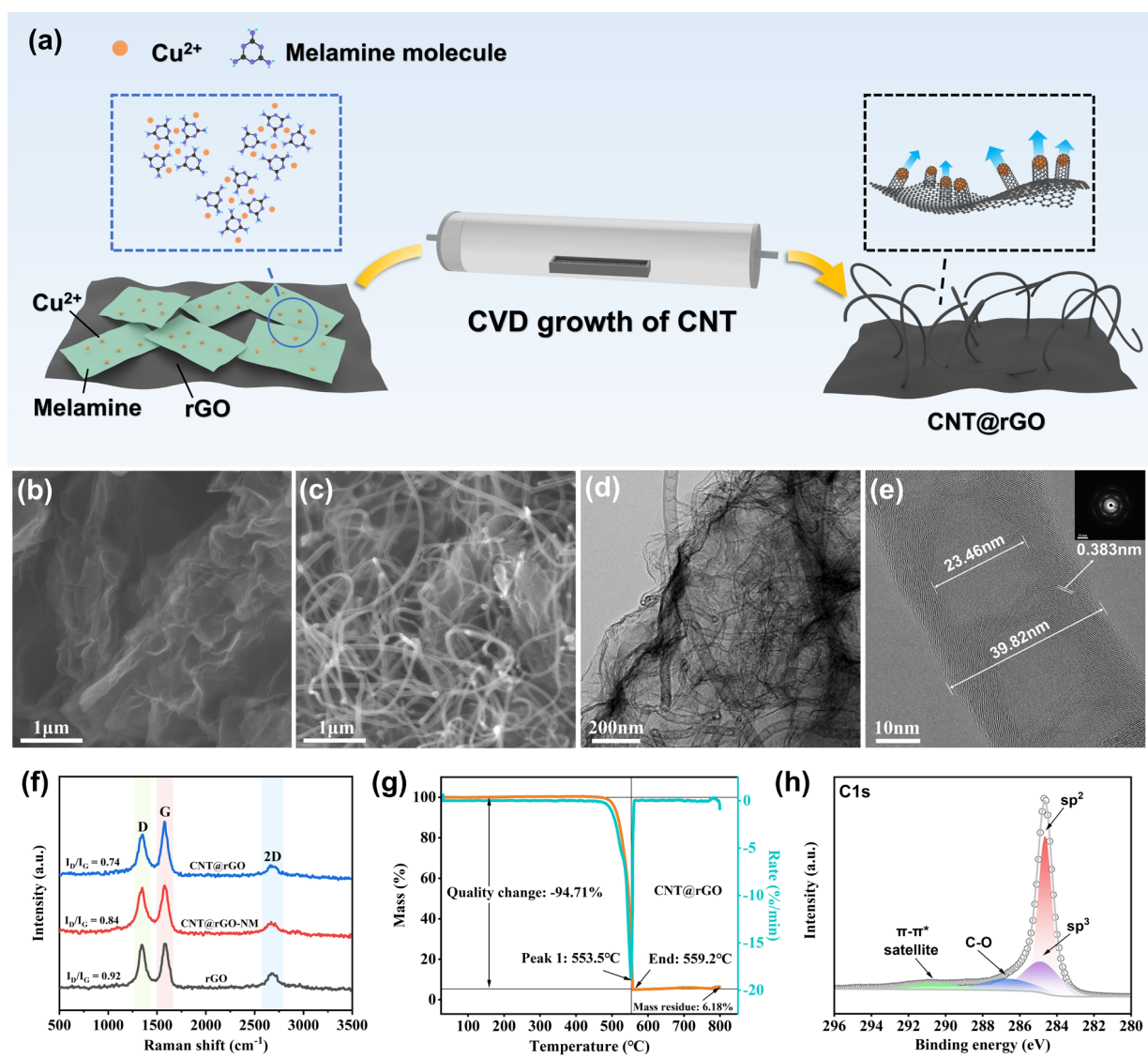

**Figure 1.** (a) Schematic of the preparation process of CNT@rGO. (b) SEM image of rGO. (c) SEM image of CNT@rGO. (d) HRTEM images of CNT@rGO. (e) HRTEM images of carbon nanotube morphology in CNT@rGO, insets are SAED images. (f) Raman spectra of rGO, CNT@rGO-NM, and CNT@rGO. (g) TG and DTG curves of CNT@rGO. (h) XPS profiles of CNT@rGO C1s.

Scanning electron microscope (SEM) was employed to obtain images of the original rGO and CNT@rGO samples; it reveals that carbon nanotubes exhibit dense and uniform growth on rGO (Figure 1b,c). Further magnification demonstrates the presence of a clear 3D structure (Figure S1a Supporting Information). To ascertain the effect of constructing a supramolecular structure on CNT growth, samples without melamine were labeled as CNT@rGO-NM. Pristine rGO exhibited a folded morphology with lamellar stacking, and carbon nanotubes grew at some locations of the CNT@rGO-NM, but in smaller amounts and with a larger difference in the morphology (Figure S2a, b Supporting Information). The favorable morphology of CNT@rGO can be attributed to the effective monatomic dispersion of melamine. A comparison of the X-ray diffraction (XRD) pattern of melamine mixed with $CuCl_2$-$2H_2O$ (Melamine-Cu) with the standard PDF card of copper reveals that the characteristic peaks of Cu are absent from the diffraction pattern of Melamine-Cu (Figure S3 Supporting Information). Copper ions in the metallic copper salt act as H-bond acceptors while melamine acts as an organic ligand to provide H-bond donors, and the two form a supramolecular structural compound through hydrogen bonding.[15] [16] The construction of the supramolecule allows the copper ions to be uniformly dispersed and anchored to the backbone, which avoids the agglomeration of ions and the formation of large ionic clusters. Concurrently, the metal particles exhibit a substantial size effect.[17] The effective dispersion of Cu by melamine facilitates the formation of nanoscale catalyst particles, which results in a measure of dissolved carbon at the CNT growth temperature. Consequently, Cu is converted from a catalytically inert to a catalytically active state, ultimately leading to CNT growth via the VSS mechanism.[18] High-resolution transmission electron microscopy (HRTEM) images demonstrate that the catalyst is located at the apex of the carbon nanotubes, with the base of the carbon tubes connected to the rGO via a junction (Figure S1b, c, and d, Supporting Information). The carbon nanotubes grown in CNT@rGO are of a dendritic nature, with a tube diameter distribution of 30-50 nm, a size determined by the formation of agglomerates of Cu nanoparticles following the pyrolysis of g-$C_3N_4$. The size of the agglomerates formed is determined by the size of the Cu nanoparticles after the pyrolysis of g-$C_3N_4$, and the dendritic morphology is due to the difference in the rate of diffusion of the carbon source on the surface of the catalyst and in the body of the large particles, which can be regarded as a series of seamlessly stacked cones. Furthermore, HRTEM observations revealed clear lattice stripes, arranged parallel to and at an angle to the growth axis, with uniform thickness on the outer wall of the tubes, and the thickness of the inner wall tapering from the bottom to the tip. In addition, selected electron diffraction spots indicate that the material possesses a well-defined crystalline structure (Figure 1d, e).

In combination with the Raman spectra (Figure 1f), a more profound comprehension of the material's chemical composition and molecular structure is attained. The distinct D and G bands at 1350 cm$^{-1}$ and 1580 cm$^{-1}$ are attributed to the disordered vibration of the structural defects and the in-plane vibration of the sp$^2$ carbon atoms, respectively. The pristine rGO $I_D/I_G$ ratio of 0.92 is attributable to defects present on the rGO lamellae, in addition to certain oxygen-containing functional groups. The decrease in $I_D/I_G$ of both CNT@rGO-NM and CNT@rGO with respect to the pristine rGO suggests that the growth of carbon nanotubes achieves a partial repair of the rGO defects. The lower degree of defects in CNT@rGO is attributed to the inhibition of amorphous carbon generation by NH$_3$ produced by the thermal decomposition of melamine and the concomitant improvement in the morphology and structure of the carbon nanotubes. Thermogravimetric analysis (TG) was performed to gain a more accurate understanding of the intrinsic composition of the material. While the pristine rGO displays a single weight loss peak at 679.3°C, corresponding to the decomposition of sp$^2$ C (Figure S4a, Supporting Information), the CNT@rGO exhibits a single weight loss peak corresponding to the decomposition of the pure phase at 553°C. The residual mass of about 6% at the conclusion of the curve is attributed to the 3% intrinsic ash residue in the pristine rGO, respectively. 3% intrinsic ash residue and 3% oxidation products of the exposed copper catalyst (Figure 1g). CNT@rGO-NM underwent two weight loss peaks corresponding to the decomposition of amorphous carbon and sp$^2$ C at 563.1°C and 563.1°C, respectively (Figure S4b, Supporting Information). The total X-ray photoelectron spectroscopy (XPS) spectrum shows the atomic component occupancy of each element (Figure S5a, Supporting Information). The high-resolution XPS spectrum of C1s can be deconvoluted into four peaks at 284.6 eV, 284.8 eV, 286.6 eV, and 290.68 eV, corresponding to the sp$^2$, sp$^3$, C-O, and π-π* chemical states, respectively. The presence of the SP$^3$ and C-O peaks is attributed to defects in the material and to partial residues of oxygen-containing functional groups (Figure S5b, Supporting Information). Finally, the ICP test results indicated the presence of Cu in the CNT@rGO sample, with a content of 0.5217%. The yield of CNT was found to be 7652.31%, and the minimal catalyst content observed resulted in the elimination of a decontamination process, thereby preserving the integrity of the material structure.

## 2.2. Characterization and calculations of carbon nanotube and rGO connection modes

Scanning transmission electron microscopy (STEM) is employed to achieve a detailed characterization of the connection between CNT and rGO in CNT@rGO structures. The secondary electron image (SEI, Figure 2a) reveals a seamless connection between CNTs and rGO. Electron energy loss spectroscopy (EELS) analysis further demonstrates distinct atomic bonding characteristics in the rGO, connection, and wall regions (Figure 2b). Specifically, rGO exhibits a higher $\sigma^*/\pi^*$ ratio compared to CNT, as the σ bonds in rGO are more aligned with the direction of momentum transfer, which is perpendicular to the beam incidence direction.[19] [20] [21] Within the CNT structure, the connection region shows a lower $\sigma^*/\pi^*$ ratio than the wall region, attributed to its orientation being closer to perpendicular to the momentum transfer direction, thereby reducing the excitation of in-plane σ-bonding electrons. The $\pi^*/\sigma^*$ intensity mapping corroborates this observation, showing a consistently lower $\sigma^*/\pi^*$ ratio across the interface region. An open-ended structural feature, indicative of covalent bonding between CNTs and rGO, is observed when the electron beam aligns with the CNT position (Figure 2c). High-resolution bright-field (BF) STEM images provide direct atomic-scale insights into the interface region (Figure 2d, e). Noise-filtered images reveal a high density of defects within the interface, as evidenced by the disruption of the periodic moiré pattern (Figure 2f). Further analysis shows that the defects are 7-membered rings. (Figure S6a-d, Supporting Information). This is consistent with the theoretical modeling, which shows the presence of 7-membered ring defects at the connection region (Figure 2g).

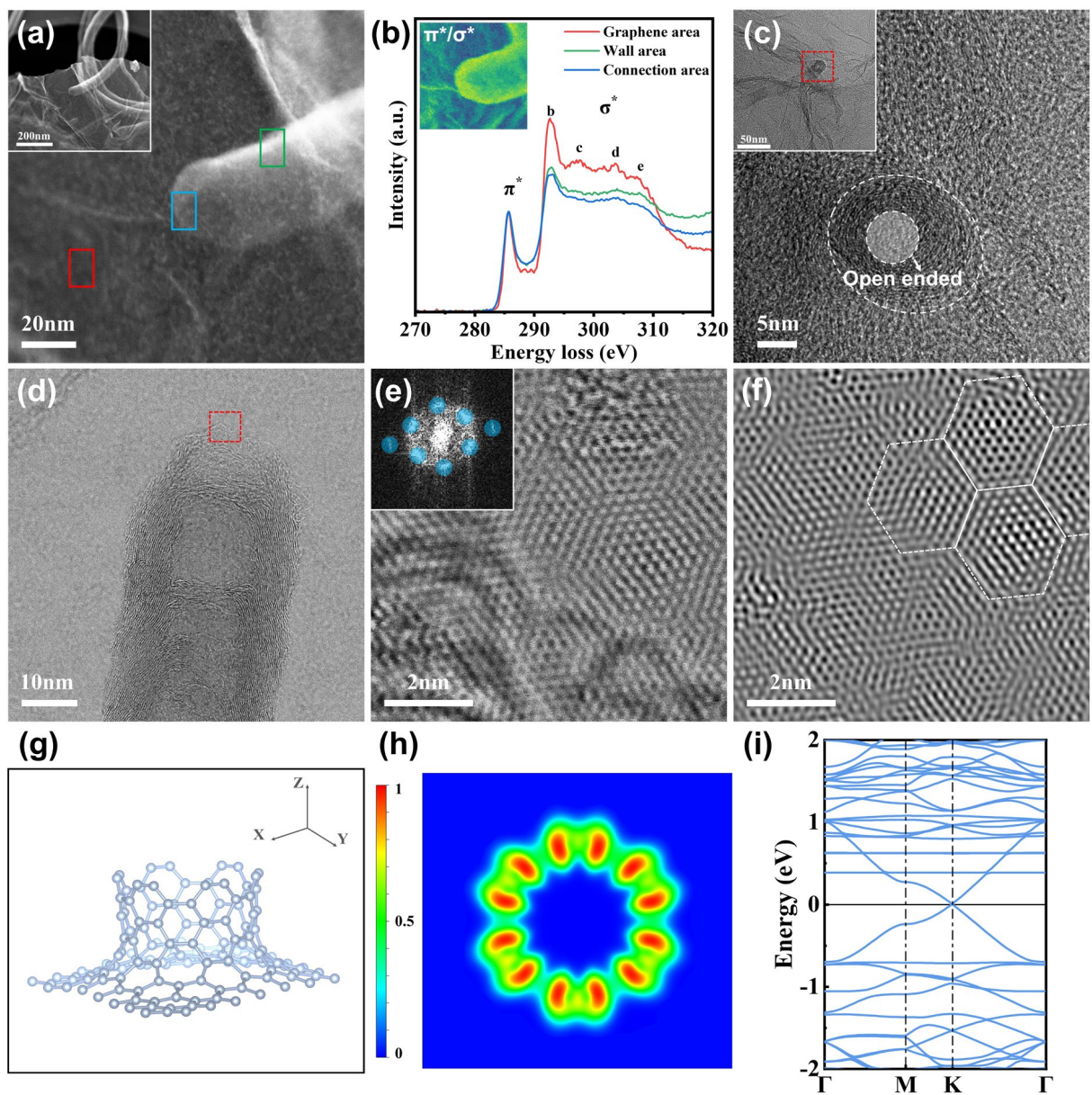

**Figure 2.** (a) SEI image of the CNT@rGO connection region, Rectangles with different colors represents different EELS acquisition regions. (b) EELS of CNT@rGO measured in different regions, inset is mapping of π*/σ* ratio intensity. (c) HRTEM image of the joint between CNT and rGO, inset is the HRTEM image of CNT on rGO slice. (d) BF STEM image of the connection area between CNT and rGO. (e) High resolution BF-STEM image showing atom positions in connection area. Insert is the FFT of the image and Blue circles shows the positions of filters. Noise of the image can be excluded by doing Inverse FFT of filter spots. (f) Inverse FFT image of (e). White dashed hexagons indicate the position of periodic moiré patterns, whose distribution is disrupted by the defects at the left bottom. (g) Modelling of CNT-graphene connection. (h) The electron localization function in the XY plane (i) Energy band structure of the CNT-graphene connection.
8

The seamless connection of graphene and nanotubes preserves the chemistry and local sp$^2$ bonding of both nanotubes and graphene. It is anticipated that this configuration will yield optimal conductance characteristics in comparison with the links that are achieved by chemical functionalization. This hypothesis is corroborated by the band structure of the system comprising graphene and a nanotube, in which the Dirac cone persists (Figure 2h, i). The covalently connected 3D nanostructures provide many unobstructed paths for electron and ion transport, allowing the materials to have good electrical properties.

## 2.3. Electrochemical properties of CNT@rGO as a conductive agent

In order to evaluate the electrochemical performance of CNT@rGO as a conductive agent for the LiFePO$_4$ positive electrode, a CR2032 button half-cell was assembled with lithium foil as the counter electrode for performance testing. To represent the profile, the name of the conductive agent was used to correspond with the given electrode. Prior to the assembly of the battery, the pole piece resistivity of the different electrodes was first obtained using the four-probe method. The results obtained are as follows: CNT@rGO has the lowest resistivity of all the other electrodes, with a value of 3.76 Ω-cm (Figure S7, Supporting Information).

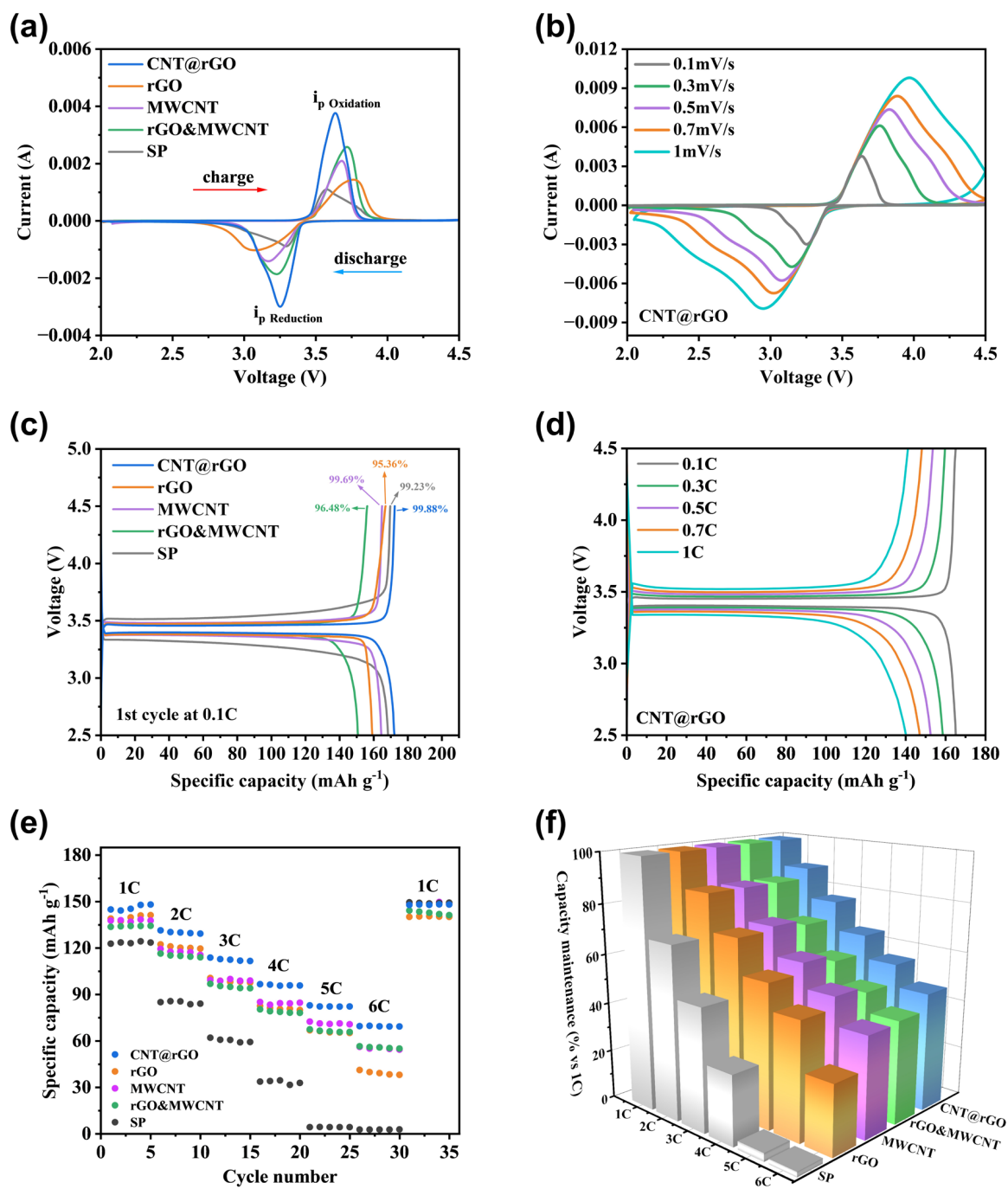

**Figure 3.** (a) CV curves of different electrodes at 0.1 mV/s sweep rate; (b) CV curves of CNT@rGO at different sweep rates; (c) Charge/discharge curves of different electrodes at 0.1C for the first cycle; (d) Charge/discharge curves of CNT@rGO at different multiplicities; (e) Multiplicity test of different electrodes at 1-6C; (f) Different electrodes at 1-6C multiplicities 3D histogram of capacity retention at relative 1C.

Following the assembly of the battery, we proceeded to conduct a series of cyclic voltammetry (CV) tests. These tests were conducted with the objective of characterizing the reaction process of the electrode during both the charging and discharging phases.

Additionally, the reversibility of the electrode reaction was determined. The peaks that appeared around 3.64 V during charging were attributed to the oxidation reaction of the LiPO4 electrode, with larger peaks corresponding to a longer charging plateau. During discharging, the peaks around 3.25 V are identified as corresponding to the reduction reaction of the electrode (Figure 3a). The absolute value of the ratio of peak current $i_{po}$ to $i_{pr}$ has been shown to reflect the degree of reversibility of the electrode, with results closer to 1 indicating a higher degree of reversibility. In addition, the peak spacing has been demonstrated to reflect the degree of polarization of the electrode. The close proximity of $|i_{po}/i_{pr}|$ of CNT@rGO to 1, in conjunction with its minimal peak spacing, signifies that the electrode is less polarized and possesses a higher degree of reversibility.[22] Furthermore, the CV tests at different sweep speeds demonstrated that the peak areas of the oxidation and reduction peaks varied to a similar extent with the increase in sweep speed (Figure 3b).[23] This finding indicates that the CNT@rGO electrode exhibited excellent reversibility performance at varying multiplicities and possessed the lowest irreversible capacity in comparison to the other electrodes within the other groups (Figure S8a, b, c, and d, Supporting Information). Subsequently, the ICE of diverse electrodes at 0.1C charging and discharging multiplicity was examined. The ICE of CNT@rGO, rGO, MWCNT, rGO&MWCNT, and SP was determined to be 99.88%, 95.36%, 99.69%, 96.48%, and 99.23%, respectively (Figure 3c). In the 0.1C to 1C multiplicity stage elevation tests, CNT@rGO exhibited a consistent and stable charging and discharging plateau, indicative of the stability and reversibility of the electrodes (Figure 3d).

Concurrently, the rate performance of the battery was evaluated. The 1-6C multiplicity test demonstrated that CNT@rGO exhibited the most stable cycling characteristics and the highest specific capacity at varying multiplicities (Figure 3e). The three-dimensional histogram of capacity retention rate visually manifested the performance discrepancy among different conductive agents (Figure 3f). In order to analyze the reasons for the difference in electrode performance, we took cross-sectional images of the pole pieces using SEM (Figure S9a, b, c, d Supporting Information); it is evident that CNT@rGO plays a pivotal role in encapsulation and bridging between $LiFePO_4$ particles, exhibiting reduced agglomeration and stacking. Conversely, agglomeration of carbon nanotubes is prevalent in MWCNT and rGO&MWCNT, resulting in an increased number of tube contact points. This can impede current transmission, thereby reducing conductivity and adversely affecting the electrodes.[24] [25] Concurrently, a precipitous decline in the capacity of rGO at 6C was observed, which may be ascribed to the fact that the structure of rGO is vulnerable to mechanical stresses during high-multiplication charging and discharging. This results in inadequate lapping between the lamellae, consequently leading to a reduction in electrical conductivity and a diminished utilization of the active material.[26] [27]

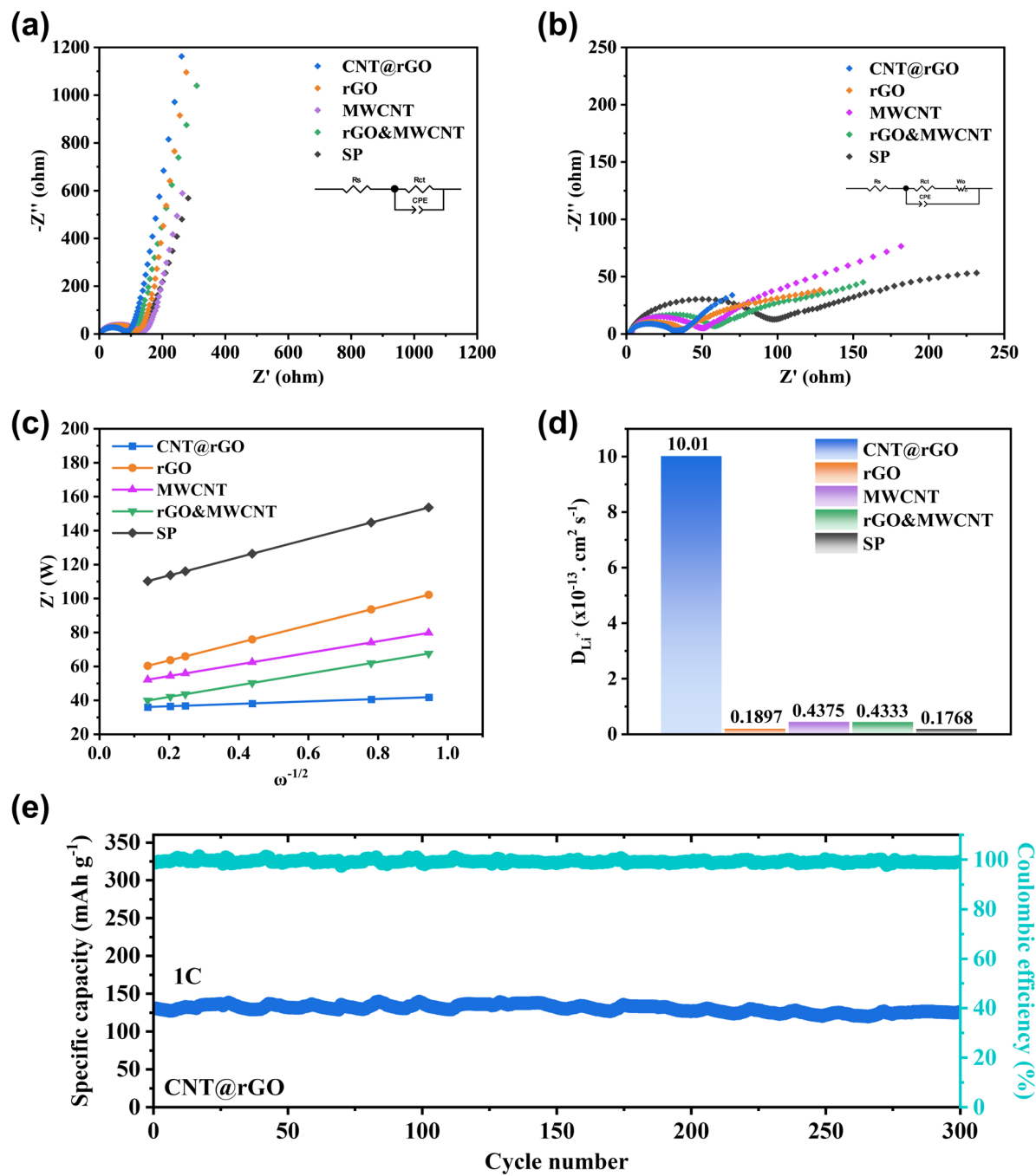

**Figure 4.** (a, b) Nyquist plots of different electrodes before and after cycling, and the inset is the fitted equivalent circuit diagram. (c) Z' versus ω-1/2; (d) Li-ion diffusion coefficients of different electrodes; (e) CNT@rGO tested at 1C for 300 cycles.

In order to conduct a more in-depth investigation into the effects of different conductive agents on the electrode electrochemical reactions, EIS tests were performed before and after the battery cycling (Figure 4a, b). In the high-frequency region, the resistance of the

CNT@rGO electrode for electron transfer, $R_{ct}$ was measured to be 79.71Ω before cycling and 29.78Ω after cycling (Figure S10a, b Supporting Information). The straight line in the low-frequency region is controlled by the diffusion of Li$^+$. In order to verify the effect of different conductive agents on the diffusion of lithium ions, we fitted and merged the impedance data to plot Z' versus the angular frequency of the low-frequency region between the inverse square root $\omega^{-1/2}$ (Figure 4c), where the slope of the straight line indicates the Warburg coefficient factor σ. The following equation is employed:[28]

$$D_{Li^+} = \frac{R^2T^2}{2A^2n^4F^4C^2\sigma^2}$$

The diffusion coefficient $D_{Li+}$ is inversely related to the Warburg coefficient factor σ, with smaller slopes corresponding to larger Li-ion diffusion coefficients. The calculation results demonstrate that CNT@rGO has the highest $D_{Li+}$, which is consistent with the performance test (Figure 4d). Finally, 300 long cycle tests were performed at a 1C charge/discharge rate, and it was found that CNT@rGO exhibited optimal cycling stability, with a residual specific capacity of 125.5 mAh g$^{-1}$ at the end of the cycle, and a capacity retention of 96.32% (Figure 4e), compared to the other electrodes (Figure S11a, b and S12 a, b Supporting Information). CNT@rGO plays a better role in the LiFePO$_4$ particles between the bridging and structural stabilization, maintaining excellent electron and ion transport properties under long cycling. When combined with the results of material characterization and electrochemical analysis, the three-dimensional conductive network architecture formed by CNT@rGO between the LiFePO$_4$ active materials significantly improves the lower electronic and ionic conductivities, enhances the structural stability of the electrodes, and significantly improves the rate performance and cycling stability of the battery.

Finally, a comparative analysis is conducted on the cost and performance of CNT@rGO conductive agent material in relation to other prevalent conductive agent materials currently available on the market (Figure S13 Supporting Information). Despite exhibiting slightly diminished performance in comparison to graphene and SWCNT, CNT@rGO demonstrates notable cost efficiency, indicating its considerable potential for practical applications in the future.

## 3. Conclusion

In summary, copper nanoparticles have been uniformly dispersed on rGO using melamine as a monatomic dispersant, and chemical vapor deposition (CVD) growth has been employed to prepare CNT@rGO with a three-dimensional nano-architecture. The carbon nanotube yield was exceptionally high at 7652.31%, with minimal catalyst residue of only 0.52%. When applied to LiFePO$_4$ cathode materials, the rate performance of batteries was superior to that of conventional conductive agents, with a capacity retention of 96.32% after 300 cycles at 1C. The results indicate that CNT@rGO, as a carbon-based conductive agent with a simple preparation process, low cost, high yield, and low catalyst residue, has high potential for application in LiFePO$_4$ cathodes for lithium-ion batteries, and it is expected to be prepared and applied on a large scale in the future.

## 4. Experimental Methods

*Commercial Materials*

LiFePO$_4$ powder was purchased from Shenzhen KejingZhida Technology Co. Carbon black (Super P) and polyvinylidene fluoride (PVDF) were obtained from Dongguan Keruder Innovative Technology Co. Multi-walled carbon nanotubes (MWCNTs) were acquired from Shandong Tianyu Nano Technology Co. Graphene oxide (GO) was synthesized from natural graphite (500 mesh, Qingdao Tengshengda Carbon Machinery Co., Ltd.) using the modified Hummers method. Reduced graphene oxide (rGO) was prepared by heat-treating graphene oxide. Melamine and copper chloride dihydrate were purchased from Shanghai Aladdin Reagent Co. All materials were used without further purification.

*Preparation of CNT@rGO*

The specific synthesis of CNT@rGO materials can be described as follows: Melamine (0.3 g) and CuCl$_2$·2H$_2$O (0.0027 g) were dispersed in anhydrous ethanol at 80°C for mixing and stirring. After the ethanol had evaporated, the obtained powder was homogenously mixed with reduced graphene oxide powder (0.1 g). Subsequently, the uniformly mixed powder was loaded into a crucible and placed in a CVD tube furnace for the growth stage of carbon nanotubes. Initially, a gas mixture of 100 sccm hydrogen and 300 sccm argon was introduced to raise the temperature to 550°C, which was maintained for 30 minutes under the same gas flow rate. Then, the hydrogen flow rate was adjusted to 5 sccm, and the argon flow rate was increased to 400 sccm, while the temperature was further raised to 850°C. Upon reaching 850°C, the temperature was held for an additional 30 minutes, during which 20 sccm ethylene, 40 sccm hydrogen, and 400 sccm argon were introduced. After natural cooling to room temperature, the final product was obtained as a black, fluffy powder.

*Calculation of yield*

$$Y = \frac{m_{mix} - m_{catalyst}X}{m_{catalyst}X} \times 100\%$$

The symbol 'Y' represents the yield. The term $m_{mix}$ denotes the mixed mass of product and residual catalyst, while $m_{catalyst}$ indicates the mass of catalyst prior to the reaction. The symbol 'X' is used to represent the relative percentage of the catalyst mass remaining after the reaction. This relative percentage is determined by analyzing the sample using Inductively Coupled Plasma Optical Emission Spectrometry (ICP-OES) after the reaction.

*Preparation of rGO&MWCNT*

The MWCNTs were compounded with rGO at a ratio of 4:6, with the resulting material designated CNT@rGO.

*Preparation of electrodes and cells*

The synthesized CNT@rGO material was used as a conductive agent for the $LiFePO_4$ positive electrode material. The powder was dissolved in N-Methyl-2-pyrrolidone (NMP) solvent and subjected to a high-pressure homogenization process for 10 minutes to obtain a CNT@rGO conductive agent slurry. Subsequently, $LiFePO_4$ (active material), polyvinylidene fluoride (PVDF), Super P, and the CNT@rGO conductive agent slurry were mixed at a weight ratio of 92:2:5:1 to prepare the electrode. The slurry was uniformly coated onto aluminum foil and dried in an air blast oven at 80°C for 1 hour, followed by vacuum drying for 12 hours. The dried electrodes were cut into discs with a diameter of 12 mm.

The assembly of the CR2032 coin cell was performed in an argon-filled glove box with $H_2O$ and $O_2$ levels maintained below 0.01 ppm. The $LiFePO_4$ electrode containing the CNT@rGO conductive agent served as the working electrode, while lithium foil was used as the counter electrode. The electrolyte consisted of 1.0 M $LiPF_6$ in a mixture of ethylene carbonate (EC) and diethyl carbonate (DEC) (v/v = 1:1), and a polypropylene (PP) membrane was used as the separator.

*Instrumental characterization*

X-ray diffraction (XRD) analysis was performed using a Rigaku Ultima IV diffractometer. Raman spectroscopy was conducted using a WITec alpha300R spectrometer, while thermogravimetric analysis (TGA) was carried out using a Netzsch TG 209 F3 thermogravimetric analyzer. X-ray photoelectron spectroscopy (XPS) measurements were obtained using a Thermo Scientific K-Alpha spectrometer. Field emission scanning electron microscopy (SEM) images were acquired using a JSM-7610F microscope at an accelerating voltage of 10 kV. Transmission electron microscopy (TEM) images were captured using a JEOL JEM-F200 microscope. Inductively coupled plasma (ICP) tests were performed using an Agilent ICP-OES 5800 spectrometer. Scanning transmission electron microscopy (STEM) images, secondary electron images (SEI), and electron energy loss spectra (EELS) were taken on an aberration-corrected JEOL GRANDARM2 scanning transmission electron microscope at 80 kV.

*Electrochemical and cell measurements*

The resistivity of the electrode was measured using a four-probe tester (Jinko Solar ST2263). Constant-current charge/discharge, rate capability, and long-cycle performance tests were conducted using a multi-channel battery tester (LAND CT2001A). Electrochemical impedance spectroscopy (EIS) measurements were carried out using an electrochemical workstation (Shanghai Chenhua CHI660E) in the frequency range of 0.1 Hz to 100 kHz.

*Calculation method*

The first-principles calculations, based on density functional theory (DFT), were carried out using the projector augmented wave (PAW) method in the VASP software package.[29] [30] [31] The electron exchange-correlation potential was approximated using the generalized gradient approximation (GGA) with the Perdew–Burke–Ernzerhof (PBE) functional.[32] The kinetic energy cutoff was set to 550 eV to enhance the plane wave basis set. The system optimization was considered complete when the total energy difference between two consecutive iterations was less than 0.001 eV, with the electronic convergence criterion set to $10^{-8}$ eV.

# Acknowledgments

This work received financial support from the National Key R&D Program of China (No. 2024YFA1207800), the National Natural Science Foundation of China (No. 22372074), the CAS Project for Young Scientists in Basic Research(YSBR-003), the Yunnan Fundamental Research Projects (202401AU070147, 202301BE070001-026), the Major Basic Research Project of Science and Technology of Yunnan (202302AG050007), Yunnan Innovation Team of Graphene Mechanism Research and Application Industrialization (202305AS350017), and Graphene Application and Engineering Research Center of Education Department of Yunnan Providence (KKPP202351001). This research benefited from resources and supports from the Electron Microscopy Center at the University of Chinese Academy of Sciences. We thank Guangdong Morion Nanotechnology Co., Ltd for providing graphene materials and technical support.

# Conflict of Interest

The authors declare no conflict of interest.

**Data Availability Statement**

The data that support the findings of this study are available from the co-responding author upon reasonable request.

ToC Figure

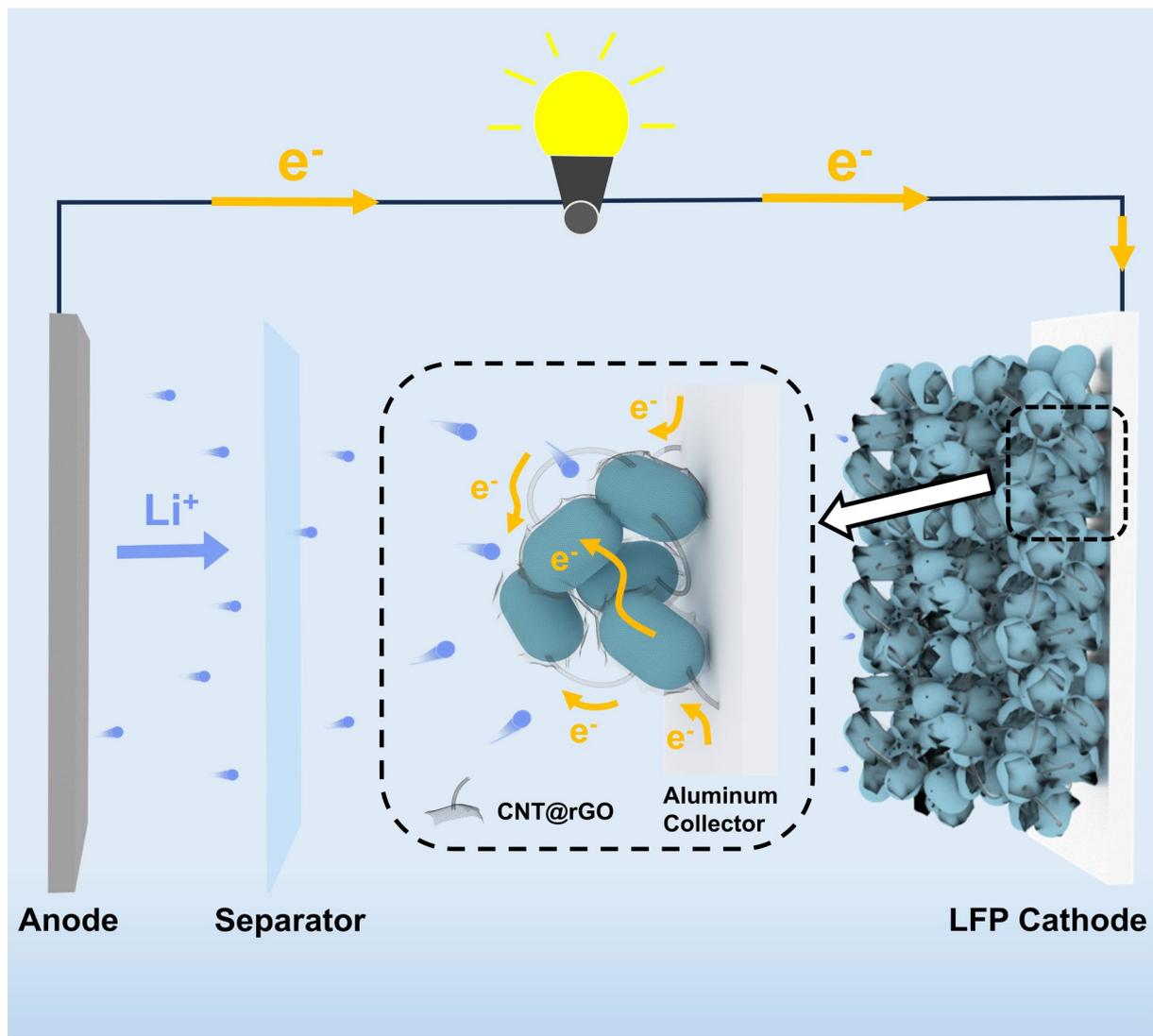